\documentclass[amsmath, amssymb, aps, prd, reprint, superscriptaddress
]{revtex4-2}

\usepackage{bm}
\usepackage{color}
\usepackage{dcolumn}
\usepackage{empheq}
\usepackage{graphicx}
\usepackage{hyperref}
\usepackage[normalem]{ulem}
\usepackage[dvipsnames]{xcolor}

\hypersetup{
    colorlinks = true,
    linkcolor  = red,
    citecolor  = blue,
}

\def\d{{\rm d}}
\newcommand{\be}{\begin{equation}}
\newcommand{\ee}{\end{equation}}
\newcommand{\bear}{\begin{eqnarray}}
\newcommand{\eear}{\end{eqnarray}}

\begin{document}

\title{Physical reinterpretation of heat capacity discontinuities for static black holes}

\author{Pedro Bargueño}
\email{pedro.bargueno@ua.es}
\affiliation{Departamento de F\'{\i}sica Aplicada, Universidad de Alicante, Campus de San Vicente del Raspeig, E-03690 Alicante, Spain}

\author{Diego Fernández-Silvestre}
\email{dfsilvestre@ubu.es}
\affiliation{Departamento de Matemáticas y Computación, Universidad de Burgos, 09001 Burgos, Spain}

\author{Juan A. Miralles}
\email{ja.miralles@ua.es}
\affiliation{Departamento de F\'{\i}sica Aplicada, Universidad de Alicante, Campus de San Vicente del Raspeig, E-03690 Alicante, Spain}

\date{\today}

\begin{abstract}
    A generic characteristic of self-gravitating systems is that they have a negative heat capacity. An important example of this behavior is given by the Schwarzschild black hole. The case of charged and rotating black holes is even more interesting since a change of sign of the specific heat takes place through an infinite discontinuity. This has been usually associated with a black hole thermodynamic phase transition appearing at the points where the heat capacity diverges, the so-called Davies points. This aspect of black hole thermodynamics has been addressed from different perspectives, motivating different interpretations since its discovery in the 1970s. In this paper, a physical reinterpretation of the heat capacity is provided for spherically symmetric and static black holes. Our analysis is partially based on a reformulation of the black hole heat capacity using the Newman--Penrose formalism. The application to the Reissner--Nordström--de Sitter black hole case reveals a clear physical interpretation of the Newman--Penrose scalars evaluated at the event horizon. This allows us to write the heat capacity as a balance of pressures defined at the horizon, in particular, a matter pressure (coming from the energy-momentum tensor) and a thermal pressure (coming from the holographic energy equipartition of the horizon). The Davies point is identified with the point where the Komar thermal energy density matches the matter pressure at the horizon. We also compare the black hole case with the case of self-gravitating objects and their corresponding thermal evolutions. We conclude that the heat capacity of black holes and self-gravitating systems can be understood qualitatively in similar terms.
\end{abstract}

\maketitle

\section{Introduction and set up of the problem}\label{SecIntro}

All the thermodynamic information of a Kerr--Newman black hole is summarized in the Smarr relation $(G = c = 1)$, \cite{Smarr:1972kt}
\begin{equation}
    M = \left[\frac{1}{4} \left(\frac{A}{4 \pi}\right) + \left(\frac{4 \pi}{A}\right) \left(J^2 + \frac{1}{4} Q^4\right) + \frac{1}{2} Q^2\right]^{1/2} \ ,
\end{equation}
where $M, J, Q$ denote the mass, angular momentum and electric charge of the black hole, respectively. The area of the event horizon $A$, as noted by Bekenstein \cite{Bekenstein1973} and Hawking \cite{Hawking:1975vcx} is proportional to the entropy $S$ associated with the event horizon, with a factor of $1/4$ ($\hbar = k_{\mathrm{B}} = 1$),
\begin{equation}
    S = \frac{1}{4} A \ .
\end{equation}
The master equation of black hole thermodynamics, $M = M \left(S, J, Q\right)$, can be written consequently as
\begin{equation}
    M = \left[\frac{S}{4 \pi} + \left(\frac{\pi}{S}\right) \left(J^2 + \frac{1}{4} Q^4\right) + \frac{1}{2} Q^2\right]^{1/2} \ .
\end{equation}
Additionally, an infinitesimal change in the black hole mass $M$ can be interpreted in terms of a first law of thermodynamics,
\begin{equation} \label{BH1}
    \delta M = T \delta S + \Omega \delta J + \Phi \delta Q \ ,
\end{equation}
where $T = (\frac{\partial M}{\partial S})_{J, Q}$, $\Omega = (\frac{\partial M}{\partial J})_{S, Q}$, and $\Phi = (\frac{\partial M}{\partial Q})_{S, J}$ are the temperature, angular velocity and electric potential of the horizon, respectively. The Smarr relation can be rewritten in terms of them simply as \cite{Smarr:1972kt}
\begin{equation} \label{Smarr}
    M = 2 T S + 2 \Omega J + \Phi Q \ .
\end{equation}

\medskip

At this point, let now us define the black hole thermal capacities generally as \cite{Davies1977}
\begin{equation}
\label{CX}
    C_X = T \left(\frac{\partial S}{\partial T}\right)_X \ ,
\end{equation}
where $X$ denotes a set of parameters that remain constant. In particular, for a Kerr--Newman black hole, the heat capacity at constant angular momentum and electric charge is \cite{Davies1977}
\begin{equation}
    C_{J, Q} = \frac{M T S^3}{\pi J^2 + \frac{\pi}{4} Q^4 - T^2 S^3} \ .
\end{equation}
One can see from this expression that the thermal capacity is $- 8 \pi M^2$ for a Schwarzschild black hole while it vanishes for the extremal Kerr--Newman case. As immediately noted by Davies, $C_{J, Q}$ suffers an infinite discontinuity. After defining $\alpha = J^2/M^4$ and $\beta = Q^2/M^2$, the so-called Davies points, where $C_{J, Q}$ diverges, are given by \cite{Davies1977}
\begin{equation} \label{eqalphabeta}
    \alpha^2 + 6 \alpha + 4 \beta - 3 = 0 \ .
\end{equation}
Specifically, for a Kerr black hole $\alpha = 2\sqrt{3} - 3$ and for a Reissner--Nordström black hole $\beta = 3/4$. Davies also explored the heat capacity discontinuities for Kerr--Newman black holes in de Sitter spacetime in a subsequent work \cite{Davies:1989ey}, generalizing the preceding results including the presence of a cosmological constant.

\medskip

At the time of Davies findings, black holes were found to have other ``critical'' points that demand physical interpretation and understanding. An example of interest concerns the seminal work by Smarr on the geometry of rotating black holes \cite{Smarr:1973zz}. In particular, the appearance of a possible instability at the Kerr event horizon near $\alpha = 1$ was proposed. What happens here is that there is a change from the electric to the magnetic part of the Weyl curvature at the event horizon near this value (the so-called Smarr point). This phenomenon is studied thoroughly, including detailed visualizations of the effect, in the works of Thorne and co-workers \cite{Zhang:2012jj}. As to Davies points, the discontinuities of the black hole heat capacity have been investigated since the discovery of Davies and different possible physical interpretations have been proposed. The same Davies considered them as phase transitions, thus initiating the study of phase transitions in black hole thermodynamics. Apart from the Davies-type, black holes were soon observed to suffer other phase transitions: the extremal phase transition \cite{Curir1981a, Curir1981b, Pavon:1988in, Pavon:1991kh} (see also \cite{Bhattacharya:2019awq}), and concerning black holes in anti-de Sitter spacetime, the Hawking--Page phase transition \cite{Hawking:1982dh}, and a van der Waals-type phase transition \cite{Dolan:2011xt, Dolan:2012jh, Kubiznak:2012wp} (see also \cite{Kubiznak:2016qmn}). Actually, the correspondence of Davies critical points with thermodynamic phase transitions was not clear and was a matter of debate. In fact, Davies himself \cite{Davies1977} commented that ``future work should be directed to understanding the full physical significance of the phase transition discussed''. In the same year as Davies's work, Hut \cite{Hut1977} discussed the fact that this infinite discontinuity may not affect the internal state of the system, as in an ordinary thermodynamic phase transition. Additionally, Sokolowski and Mazur \cite{Sokolowski:1980uva} claimed that the heat capacity change of sign and infinite discontinuity have a geometric cause (determined by the embedding of an event horizon into a black hole spacetime). They commented that ``the simplest resolution of the problem of the black hole phase transition would therefore be that the phenomenon is due merely to the fact that the temperature is not a monotonic function of the entropy'', concluding that ``a full elucidation of the problem is still missing''. After some years, Kaburaki and co-workers \cite{Kaburaki:1993ah, Katz:1993up, Kaburaki1994, Kaburaki1996} argued that the Davies points are not critical points but just turning points indicating a change of stability of the system. Alternative geometrical approaches to understanding the heat capacity discontinuities were proposed by Cai and co-workers \cite{Shen:2005nu}, showing that for the Reissner--Nordström and Kerr black holes, a particular thermodynamic geometry, coined Ruppeiner geometry \cite{Ruppeiner:2008kd}, was curved and the scalar curvature turned out to approach negative infinity at the discontinuities. These ideas led Quevedo and co-workers to introduce, within a geometrothermodynamics approach, a Legendre-invariant metric in the space of thermodynamic equilibrium states containing all the information about the thermodynamics of black holes \cite{Alvarez:2008wa}. The curvature of this thermodynamic metric for Kerr--Newman black holes turned out to be singular at the Davies points. A connection between the dynamical and thermodynamical properties of black holes has also been suggested, in particular, between the Davies points and the quasinormal modes of charged black holes, both in asymptotically flat \cite{Jing:2008an, Berti:2008xu} and asymptotically de Sitter spacetimes \cite{Wei:2019jve}. On the other hand, a topological approach has been considered more recently to determine and explore black hole properties without resorting to concrete structure details and/or dynamical field equations. These arguments have been applied to the Davies points \cite{Bhattacharya:2024bjp, Hazarika:2024imk}, giving them a topological interpretation.

\medskip

In our opinion, despite the different aforementioned interesting proposals to try to interpret the black hole heat capacity discontinuities, we have not found a fully satisfying and simple physical explanation for them. It is then our aim to present an alternative approach to this problem based on a reformulation of black hole heat capacities in terms of Newman--Penrose scalars and a comparison of this alternative formulation with the heat capacities of self-gravitating systems.

\medskip

This manuscript is organized as follows: In Sec.~\ref{SecNP}, heat capacities for spherically symmetric black holes are rewritten in
terms of the Newman--Penrose scalars. After applying this formalism to the Reissner--Nordström--de Sitter case in Sec.~\ref{SecRN},
a physically transparent criteria for the occurrence of the Davies points is revealed. Finally, the heat capacity of nondegenerate and degenerate self-gravitating objects is described in Sec.~\ref{SecJA} in order to compare the essential physics with that of black holes. We conclude with some final remarks in Sec.~\ref{Conclusions}.

\section{Newman--Penrose reformulation of black hole heat capacity} \label{SecNP}

In the following section, black hole heat capacities will be expressed in terms of the Newman--Penrose scalars. The reason behind doing that will be transparent when the Reissner--Nordström--de Sitter case, analyzed in light of the mentioned reformulation, is compared to standard models of star evolution. The aim is to give a clear physical interpretation of the divergence of heat capacities for spherically symmetric and static black holes.

\medskip

Let us begin by considering a spherically symmetric and static spacetime, with line element
\begin{equation}\label{metric}
    ds^2 = f(r) dt^2 - \frac{dr^2}{f(r)} - r^2 d\Omega^2 \ ,
\end{equation}
with $d\Omega^2 = d\theta^2 + \sin^2 \theta d\phi^2$ the metric of the two-sphere.

\medskip

On the one hand, let us briefly review the concept of Komar energy \cite{Komar:1958wp} in general stationary spacetimes. The Komar energy is the associated conserved charge with the spacetime isometry generated by the timelike Killing vector field. This quantity, when 
calculated at infinity, is interpreted as the total energy of the spacetime. The Komar energy $E_K$ is given by the following surface integral:
\begin{equation}
    E_K = \frac{1}{4 \pi} \int_{\partial \Sigma} d^2x \sqrt{\gamma} \, n_\mu \sigma_\nu \nabla^\mu K^\nu \ .
\end{equation}
In the above expression, also called the Komar integral, $\partial \Sigma$ is the boundary of a spacelike hypersurface $\Sigma$. The unit normal vectors of $\Sigma$ and $\partial \Sigma$ are, respectively, $n^\mu$ and $\sigma^\mu$, and $\gamma$ is the induced metric on $\partial \Sigma$. Finally, $K^\mu$ is the timelike Killing vector. One can show that, for the case of Eq.~\eqref{metric}, the Komar energy is simply
\begin{equation} \label{EKomar}
    E_K = \frac{1}{2} r^2 f'(r) \ .
\end{equation}
If the Komar energy is evaluated at an event horizon, the result is
\begin{equation} \label{EKomarH}
    E_{KH} = \frac{1}{2} \frac{A}{l_P^2} T = \frac{1}{2} N T \ ,
\end{equation}
where $A$ is the area of the horizon, $T$ is the associated Hawking temperature, and $N$ is the number of degrees of freedom of the horizon. The above equation describes a law of equipartition of energy among the degrees of freedom living on the horizon, with the horizon understood as tessellated in elements of area $l_P^2$. Note that we have (temporarily) introduced the Planck length $l_P$ in order to have a clear interpretation of the degrees of freedom. This is the so-called holographic energy equipartition for static spacetimes \cite{Padmanabhan:2009vy}.

\medskip

On the other hand, following Penrose and Rindler's conventions \cite{Penrose1984}, let us introduce the following null tetrad:
\begin{equation}
    \begin{aligned}
        l^\mu &= \left(\frac{1}{\sqrt{2 f(r)}}, - \sqrt{\frac{f(r)}{2}}, 0, 0\right) \ , \\
        n^\mu &= \left(\frac{1}{\sqrt{2 f(r)}}, \sqrt{\frac{f(r)}{2}}, 0, 0\right)\ , \\
        m^\mu &= \left(0, 0, - \frac{1}{\sqrt{2} r}, \frac{\textrm{i} \csc \theta}{\sqrt{2} r}\right) \ ,
    \end{aligned}
\end{equation}
where $l_\mu n^\mu = 1$ and $m_\mu \bar{m}^\mu = - 1$, with the bar denoting complex conjugation. In this case, the only nonvanishing Newman--Penrose scalars are
\begin{equation}
    \begin{aligned}
        \Psi_2 &= C_{\mu \nu \rho \sigma} l^\mu m^\nu \bar{m}^\rho n^\sigma \ , \\
        \Phi_{11} &= - \frac{1}{2} R_{\mu \nu} l^\mu n^\nu + 3 \Lambda \ , \\
        \Lambda &= \frac{R}{24} \ , \\
    \end{aligned}
\end{equation}
where $C_{\mu \nu \rho \sigma}$ and $R_{\mu \nu}$ stand for the Weyl and Ricci tensors, respectively, and $R = g^{\mu \nu} R_{\mu \nu}$ is the Ricci curvature scalar. Let us note that, interestingly, for the geometries given in Eq.~\eqref{metric}, it is possible to solve for the metric function $f(r)$ and its derivatives in terms of the Newman--Penrose scalars, obtaining
\begin{equation} \label{fs}
    \begin{aligned}
        f(r) &= 1 + 2 r^2 \left(\Psi_2 - \Phi_{11} - \Lambda\right) \ , \\
        f'(r) &= -  2 r \left(\Psi_2 + 2 \Lambda\right) \ , \\
        f''(r) &= 4 \left(\Psi_2 + \Phi_{11} - \Lambda\right) \ . 
    \end{aligned}
\end{equation}
The Komar energy \eqref{EKomar} can consequently be expressed as a function of the Newman--Penrose scalars,
\begin{equation} \label{EKomarNP}
E_K = - r^3 (\Psi_2 + 2 \Lambda) \ .
\end{equation}
Additionally, taking into account the second equation in Eq.~\eqref{fs}, the temperature in terms of the Newman--Penrose scalars evaluated at the horizon is
\begin{equation} \label{TNP}
    T = - \frac{1}{2 \pi} r_H \left(\Psi_{2H} + 2 \Lambda_{H}\right) \ .
\end{equation}
The above equation can also be expressed as a function of the horizon entropy, considering that $S = \pi r_{H}^2$ for spacetimes with spherical symmetry. Notice that Eq.~\eqref{EKomarH} is correctly reproduced when Eq.~\eqref{EKomarNP} is evaluated at the event horizon.

\medskip

Finally, the heat capacity can be computed using Eqs.~\eqref{CX} and \eqref{TNP}, where it is understood that $\Psi_{2H} = \Psi_{2H} (X, S)$.  After a short calculation, we obtain
\begin{equation} \label{EqGeneral}
    C_X = 2 S \, \frac{\Psi_{2H} + 2 \Lambda_{H}}{\Psi_{2H} + 2 \Lambda_{H} + 2 S \left(\frac{\partial \Psi_{2H}}{\partial S}\right)_X} \ .
\end{equation}

\section{Application to the Reissner--Nordström--de Sitter spacetime} \label{SecRN}

As an application of Eq.~\eqref{EqGeneral}, let us now consider a Reissner--Nordström black hole embedded in a de Sitter spacetime. The line element is given by 
Eq.~\eqref{metric} with
\begin{equation}
    f(r) = 1 - \frac{2 M}{r} + \frac{Q^2}{r^2} - \frac{\lambda}{3} r^2 \ ,
\end{equation}
with the (positive) cosmological constant denoted as $\lambda$. The nonvanishing Newman-Penrose scalars are
\begin{equation}\label{NPRN}
    \begin{aligned}
        \Psi_2 &= - \frac{M r- Q^2}{r^4} \ , \\
        \Phi_{11} &= \frac{Q^2}{2 r^4} \ , \\
        \Lambda &= \frac{\lambda}{6} \ .
    \end{aligned}
\end{equation}
In order to efficiently employ Eq.~\eqref{EqGeneral}, the Penrose--Rindler equation \cite{Bargueno2023} is especially important. It states that
\begin{equation}
    - \Psi_{2H} + \Phi_{11H} + \Lambda_H = \frac{K_G}{2} \ ,
\end{equation}
where $K_G = \frac{1}{r_{H}^2}$ is the Gaussian curvature of the horizon, which is a two-sphere. Note that the Penrose--Rindler equation is nothing but the first equation in Eq.~\eqref{fs} evaluated at the event horizon. In the case under consideration, $\Lambda$ is proportional to the cosmological constant, with a factor of $1/6$, so $\Lambda_H = \Lambda$.

\medskip

The aim is now to calculate the heat capacity at constant electric charge, so taking into account that $\Phi_{11H} = \frac{\pi^2 Q^2}{2 S^2}$, by means of the Penrose--Rindler equation we have that $\Psi_{2H} = \frac{\pi^2 Q^2}{2 S^2} - \frac{\pi}{2 S} + \Lambda$, and the derivative in Eq.~\eqref{EqGeneral} gives
\begin{equation}
    \left(\frac{\partial \Psi_{2H}}{\partial S}\right)_Q = - \frac{1}{S} (\Psi_{2H} + \Phi_{11H} - \Lambda) \ .
\end{equation}
Finally, the heat capacity at constant $Q$ reads
\begin{equation} \label{HCQNP}
    C_Q = - \frac{2 \pi}{K_G} \frac{\Psi_{2H} + 2 \Lambda}{\Psi_{2H} + 2 \Phi_{11H} - 4 \Lambda} \ .
\end{equation}
with the Davies points showing up when
\begin{equation}
    \Psi_{2H} + 2 \Phi_{11H} - 4 \Lambda = 0 \ .
\end{equation}

\subsection{Physical interpretation}

We are now ready to reanalyze the heat capacity, with its infinite discontinuity. The physical interpretation provided in this section will become clearer in the following section when compared with the heat capacity of self-gravitating objects with matter. One can check, in the first place, that Eq.~\eqref{HCQNP} correctly reproduces the results in \cite{Davies:1989ey} when Eq.~\eqref{NPRN} is explicitly introduced and evaluated at the horizon. We will now see that, apart from the factor of the Gaussian curvature of the horizon, the rest of Eq.~\eqref{HCQNP} is especially amenable to an interesting physical interpretation.

\medskip

Let us rewrite Eq.~\eqref{HCQNP} as
\begin{equation} \label{HCQNP2}
    C_Q = - \frac{2 \pi}{K_G} \frac{\Psi_{2H} + 2 \Lambda}{\Psi_{2H} + 2 \Lambda + 2 \left(\Phi_{11H} - 3 \Lambda\right)} \ .
\end{equation}
Notice now that for a Reissner--Nordström--de Sitter black hole, the pressure associated with the Maxwell source term on the black hole horizon is given by
\begin{equation} \label{PQ}
    \frac{\Phi_{11H}}{4\pi} = P_Q = \frac{Q^2}{8 \pi r_H^4} \ ,
\end{equation}
while 
\begin{equation} \label{Plambda}
- \frac{3 \Lambda}{4\pi} = P_\lambda = - \frac{\lambda}{8\pi}
\end{equation}
is the pressure associated with the cosmological constant. Let us highlight that the electrostatic pressure introduced in Eq.~\eqref{PQ} is simply the pressure exerted by a spherical conductor of radius $r_H$, carrying an electric charge $Q$, on its surface (in units such that $4 \pi \epsilon_0 = 1$, where $\epsilon_0$ is the electric constant). In this context, this pressure is then interpreted as the pressure that the black hole electric charge, more concretely, the resultant electric field, exerts on the horizon. What is more, a thermal pressure at the black hole horizon can be defined as
\begin{equation} \label{Pth}
    - \frac{1}{4 \pi} \left(\Psi_{2H} + 2\Lambda\right) = P_{\mathrm{th}} = \frac{\bar{N} T}{V} \ ,
\end{equation}
where we have introduced the areal volume, $V = \frac{4}{3} \pi r_H^3$. In the case of spherical symmetry, this volume coincides exactly with the thermodynamic volume in the so-called extended phase space approach to black hole thermodynamics \cite{Kubiznak:2016qmn}. Alternatively, the radial Einstein equation for spherically symmetric spacetimes can be interestingly interpreted as an equation of state $P = P (T, V)$ when the areal volume $V$ is assumed to coincide with the thermodynamic volume and $P = - T_r^r$, where $T_{\mu \nu}$ is the energy-momentum tensor \cite{Padmanabhan:2002sha}. In this context, a black hole can be roughly thought of as an ensemble of $\bar{N} = \frac{N}{6}$ ``particles'' \cite{Altamirano:2014tva} inside a volume $V = \frac{4}{3}\pi r_H^3$ at equilibrium at the Hawking temperature $T$ and supported by the matter pressure $P$, namely, the sum of pressures appearing in the energy-momentum tensor, including a possible cosmological constant. This allows us to write Eq.~\eqref{HCQNP2} as
\begin{equation} \label{HCNP3}
    C_Q = - \frac{2 \pi}{K_G} \frac{1}{1 - 2 \frac{P_Q + P_{\lambda}}{P_{\mathrm{th}}}} \ .
\end{equation}
The heat capacity therefore diverges when
\begin{equation}
    P_{\mathrm{th}} = 2 \left(P_Q + P_{\lambda}\right)
\end{equation}
or, equivalently, when
\begin{equation}
    \frac{\bar{E}_{KH}}{V} = P_Q + P_\lambda \ ,
\end{equation}
where $\bar{E}_{KH}$ denotes the Komar energy \eqref{EKomarH} for $\bar{N} = \frac{N}{6}$ degrees of freedom. We can then conclude that the Davies point occurs precisely when the density of the Komar (thermal) energy $\bar{E}_{KH}$, given by the holographic energy equipartition of the horizon, matches with the sum of pressures appearing in the energy-momentum tensor (nonthermal), $P_Q + P_\lambda$.

\medskip

Once the physical interpretation of the heat capacity of the Reissner--Nordström--de Sitter black hole (as associated with a balance of pressures at the horizon) and the corresponding Davies points (as a certain equilibrium at the horizon) has been provided thanks to the Newman--Penrose formalism, in the following section the heat capacity of self-gravitating objects will be studied in terms of the thermal evolution of such systems in order to compare it with the black hole case.

\section{Heat capacity of self-gravitating objects} \label{SecJA}
The basic mechanism leading to star formation consists of the fact that nondegenerate self-gravitating objects without nuclear energy sources have negative heat capacity \cite{LBLB1977}. In this case, the evolution of such objects is thermally unstable in the sense that the more energy is lost, the higher the temperature is. This process ends when the central temperature reaches a sufficiently high value to initiate nuclear burning of the main constituent element, namely, hydrogen. Along with the increase in temperature, the density also increases due to the contraction of the object. Actually, the conservation of energy establishes that the increase in internal energy together with the energy losses must be compensated by the decrease in gravitational energy. The contraction of the object can affect the behavior of matter by changing the relation between pressure, temperature, and density, the so-called equation of state. At low densities, the pressure is proportional to the temperature (nondegenerate gas), but at high densities the pressure is almost independent of the temperature (degenerate gas). It is precisely in this transition when the heat capacity of the self-gravitating object turns out to change sign and the temperature decreases with the emission of energy, instead of increasing, and the self-gravitating object becomes thermally stable and begins to cool. This prevents the formation of stars (self-gravitating systems burning hydrogen) if the mass is not high enough ($0.08 M_\odot$) (see, for instance, \cite{KWW2012}). 

\medskip

\begin{figure}[ht]
    \centering
    \includegraphics[scale=0.9]{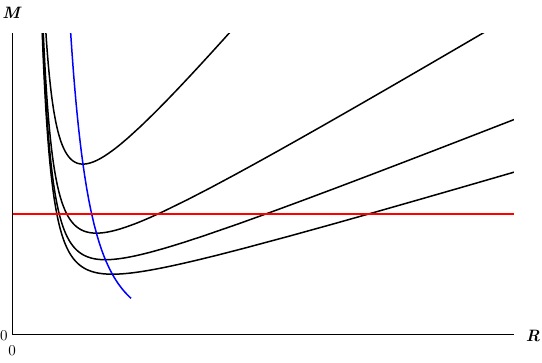}
    \caption{Isothermal curves for a self-gravitating object in the $R$-$M$ plane. See text for details.}
    \label{Fig_Iso}
\end{figure}

Let us build the simplest possible model that captures the essential features of the outlined process. This model will allow us to emphasize the similar behavior between the thermal evolution of black holes and self-gravitating objects. We write the total energy of the object $E$ as the sum of the internal energy $U$ and the gravitational energy $\Omega$,
\begin{equation}
    E = U + \Omega \ .
\end{equation}
On the other hand, the hydrostatic equilibrium equations 
\begin{equation}
    \begin{aligned}
        \frac{\d P}{\d r} &= - G \frac{m}{r^2} \rho \\
        \frac{\d m}{\d r} &= 4 \pi r^2 \rho \ ,
    \end{aligned}
\end{equation}
lead to the virial theorem which, for an ideal monatomic gas, is written as
\begin{equation}
\Omega = - 2 U \ ,
\end{equation}
so that the total energy of the object reads
\begin{equation}
E = - U = \frac{1}{2} \Omega \ .
\end{equation}
A release of energy produces then two effects, an increase in internal energy for the same amount of the energy released and a decrease in gravitational energy for twice the energy released, namely,
\begin{equation}
\delta E = - \delta U = \frac{1}{2} \delta \Omega \ .\label{dvirial}
\end{equation}
We have to keep in mind that even if the gas is ideal, an increase in the internal energy does not imply that the temperature has to increase as well. Now, gravity is causing the density to increase by doing work on the system. As the gas approaches degeneracy, the internal energy and the pressure become more and more dependent on the density than on the temperature. Let us show this by considering the following simple relation between the internal energy, the density $\rho$, and the temperature $T$:
\begin{equation}
    \begin{aligned}
        U &= \frac{3}{2} P V = 2 \pi R^3 \left(P_{\rm th} + P_{\rm deg}\right) \\
        &= 2 \pi R^3 \left(\frac{\rho}{\mu m_u} k_B T + K \rho^{5/3} \right) \ ,
    \end{aligned}
\end{equation}       
where $P$ denotes the total pressure, $V$ the volume, $m_u$ the atomic mass unit, and $\mu$ the mean molecular weight, while $k_B$ is the Boltzmann constant and $K$ is a different constant. In the above expression, we have assumed a monatomic gas and we have written the total pressure as the sum of the thermal pressure and the degeneracy pressure. Although this expression may not be precise, for our purposes it is sufficient. Let us proceed further by eliminating the density in terms of the mass $M$ and the radius $R$ of the object, using the average density value $\frac{3 M}{4 \pi R^3}$. We get the relation
\begin{equation}
U = \frac{3 k_B}{\mu m_u} M T + \frac{3^{5/3}}{(4\pi)^{2/3}} K \frac{M^{5/3}}{R^2} \ .
\end{equation}
We can now use the virial theorem to relate the internal energy and the gravitational energy $\Omega \sim - G \frac{M^2}{R}$, finally obtaining the temperature as a function of the mass and the radius,
\begin{equation}
T = \frac{\mu m_u}{3k_B} \left(G\frac{M}{R} - 3\left(\frac{3}{4\pi}\right)^{2/3} K \frac{M^{2/3}}{R^2}\right) \ .
\end{equation}
Figure \ref{Fig_Iso} shows the isothermal curves (black lines) in the plane $R$-$M$. The highest temperatures correspond to the upper curves. The blue line joins the minima of the isothermal curves. The red line shows the evolution of a self-gravitating object from right to left. Its mass remains constant throughout evolution (losing energy) while contraction shows up as a decreasing radius. The temperature always increases until it reaches a minimum of an isothermal curve, namely the point where the red line intersects the blue line. After that point, the temperature begins to decrease.

\medskip

\begin{figure}[ht]
    \centering
    \includegraphics[scale=0.9]{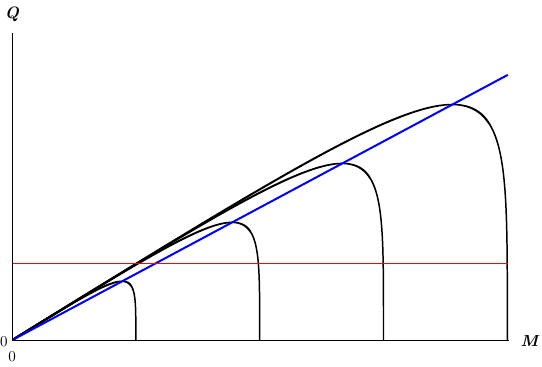}
    \caption{Isothermal curves for the Reissner--Nordström black hole in the $M$-$Q$ plane. See text for details.}
    \label{Fig_Iso_RN}
\end{figure}

A similar situation takes place in the quite different context of the thermal evolution of black holes. Figure \ref{Fig_Iso_RN} shows the isothermal curves for a Reissner--Nordström black hole (black lines) in the plane $M$-$Q$. In this case, the highest temperatures correspond to the lower curves. The blue line joins the maxima of the isothermal curves. The red line describes the evolution of the black hole, from right to left, during the loss of energy (mass) due to Hawking radiation, while the electric charge remains constant. The temperature of the Reissner--Nordström black hole increases (negative heat capacity) until reaching the maximum of an isothermal curve. The temperature begins to decrease (positive heat capacity) once the mass becomes less than that corresponding to the Davies point, i.e., $\frac{2}{\sqrt{3}}Q$, eventually reaching zero temperature (extremal Reissner--Nordström black hole). This analysis follows from considering the Reissner--Nordström black hole evolution under Hawking radiation at constant electric charge. In the evaporative evolution, the loss of charge should also be taken into account. It was shown in \cite{Hiscock:1990ex} that the thermal evolution of black holes can approximately be divided into two regions:  a ``charge dissipation zone'' and a ``mass dissipation zone.'' A Reissner--Nordström black hole with sufficiently large mass always starts evolving in the mass dissipation zone, where the evolution is indeed very well approximated by the loss of mass with the charge remaining constant. It was also shown in \cite{Hiscock:1990ex} that a black hole with a mass below a certain value will never have a change of sign in the heat capacity, so it will never have a Davies point. This fact prevents the final step of the thermal evolution from being the extremal limit. The reader is referred to Ref.~\cite{Hiscock:1990ex} for a complete description of the evaporative evolution of Reissner--Nordström black holes.

\medskip

As before, we build a very simple model that captures the essential features of the heat capacity of a self-gravitating object and estimate the point where it changes sign. One can check more sophisticated models that take into account the pressure distribution across the star in the context of the cooling of white dwarfs in, for instance, \cite{KC1990}. The emission of energy while maintaining hydrostatic equilibrium is a process that can be divided into two thermodynamic processes: the heat release (corresponding to the energy losses) and the work done by gravity due to the contraction of the star. The first law of thermodynamics gives 
\begin{equation}
\delta U = \delta Q - \delta W = \delta E - P \delta V \ ,
\end{equation}
and, from Eq.~\eqref{dvirial}, $\delta Q = \delta E = - \delta U$ and $\delta W \equiv P \delta V = - 2 \delta U$. Let us now estimate the heat capacity $\mathcal{C}$ of the self-gravitating object, which is defined as 
\begin{equation}
\mathcal{C} \equiv \frac{\delta E}{\delta T} = - \frac{\delta U}{\delta T} \ .
\end{equation}
If we consider the internal energy $U$ as a function of $T$ and $V$, 
\begin{equation}
    \begin{aligned}
        \delta U &= \left(\frac{\partial U}{\partial T}\right)_V \delta T + \left(\frac{\partial U}{\partial V}\right)_T \delta V \\
        &= C_V\delta T - \frac{2}{P} \left(\frac{\partial U}{\partial V}\right)_T \delta U \ ,
    \end{aligned}
\end{equation} 
where $C_V$ is the heat capacity at constant volume. The above expression leads to
\begin{equation}
    \mathcal{C} = - \frac{C_V}{1 + \frac{2}{P}\left(\frac{\partial U}{\partial V}\right)_T} \ .
\end{equation}
Using the thermodynamic relations
\begin{equation}
    \begin{aligned}
        \left(\frac{\partial U}{\partial V}\right)_T &= T \left(\frac{\partial S}{\partial V}\right)_T - P \ , \\
        \left(\frac{\partial S}{\partial V}\right)_T &= \left(\frac{\partial P}{\partial T}\right)_V \ ,
    \end{aligned}
\end{equation}
yields
\begin{equation}
    \mathcal{C} = - \frac{C_V}{2\left(\frac{\partial \ln P}{\partial \ln T}\right)_V - 1} \ . 
\end{equation}
If we decompose the pressure into two terms  explicitly as $P = P_{\rm th} + P_{\rm deg}$, with $P_{\rm th}$ the thermal pressure and $P_{\rm deg}$ the degeneracy pressure (which is independent of the temperature), we end up with the following expression:
\begin{equation}
    \mathcal{C} = - \frac{C_V}{\frac{2P_{\rm th}}{P_{\rm th} + P_{\rm deg}} \left(\frac{\partial \ln P_{\rm th}}{\partial\ln T}\right)_V - 1} \ . \label{C}
\end{equation}
If the thermal pressure dominates with respect to the degeneracy pressure, the denominator of this equation becomes $+1$ and the heat capacity of the self-gravitating object is negative, its absolute value being the heat capacity at constant volume.

The star increases its temperature as it emits energy through the surface. On the other hand, contraction brings the object closer and closer to degeneracy and, eventually, the degeneracy pressure starts dominating. The heat capacity then becomes positive and equal to the heat capacity at constant volume when the thermal pressure is negligible. This change of sign of the heat capacity characterizing this transition occurs when the denominator of the right-hand side of Eq.~\eqref{C} vanishes, namely, when 
\begin{equation}
    \frac{2P_{\rm th}}{P_{\rm th} + P_{\rm deg}} \left(\frac{\partial \ln P_{\rm th}}{\partial\ln T}\right)_V = 1 \ .
\end{equation}

\medskip

Again, let us compare with the black hole scenario. In Sec.~\ref{SecRN}, we have introduced the thermal pressure associated with the Komar energy density [see Eq.~\eqref{Pth}], as well as the sum of pressures appearing in the energy-momentum tensor, related to the electric charge and the cosmological constant [see Eqs.~\eqref{PQ} and \eqref{Plambda}]. If we consider the Reissner--Nordström black hole ($\lambda = 0$) the discontinuity in the heat capacity occurs when the thermal pressure is twice the electrostatic pressure. If the thermal pressure dominates with respect to the electrostatic pressure, the heat capacity is negative. The black hole increases its temperature as it emits energy (mass) via Hawking radiation with the charge remaining constant. This causes the black hole to contract. That is, the area of the event horizon decreases during its evolution at constant charge. As a consequence, the contribution of the electrostatic pressure at the horizon will be more and more important. The heat capacity becomes positive when the electrostatic pressure dominates over the thermal pressure. This change of sign of the heat capacity characterizing this transition occurs precisely at the Davis point.

\section{Final remarks} \label{Conclusions}

In this work, the Newman--Penrose formalism has been considered to derive a general expression for the heat capacity of spherically symmetric and static black holes characterized by a single function in the metric. We have been able to identify in this way the occurrence of Davies points with a certain combination of the Newman--Penrose scalars of this particular solution, which can be ascribed with a thermodynamic and physically transparent interpretation. We have shown specifically that the heat capacity of the Reissner--Nordström--de Sitter black hole can be physically understood as a particular balance of pressures at the horizon. The corresponding divergence turns out to appear when the Komar energy density is equal to the sum of pressures appearing in the energy-momentum tensor, which includes the pressures associated with the Maxwell source term of Einstein's equations as well as the cosmological constant. We have finally compared black holes and self-gravitating systems by computing the heat capacity of objects with matter and contrasting their thermal evolutions with the help of the isothermal curves. We have checked that the behavior of the heat capacity is qualitatively similar in the case of degenerate stars, with the degeneracy pressure playing the role of the sum of pressures appearing in the energy-momentum tensor. We can conclude then that the physical nature of the black hole heat capacity divergences is, in this sense, not so different from that of self-gravitating systems.

\medskip

The next step now is to try to account for rotation as well. One can conjecture that rotating black holes could present an additional pressure term associated with the rotation at the horizon. In this case, however, the Weyl scalar $\Psi_2$ has both real (electric) and imaginary (magnetic) parts, with the latter related to the rotation. As a consequence, the inclusion of rotation is not obvious. This case is currently under investigation and details are postponed for future work.

\medskip

We would like to mention that the analysis of this article is completely general, as far as static black holes are concerned. We have only considered black hole spacetimes with spherical horizons, but our discussion is easily extended to nonspherical horizons, such as hyperbolic or planar horizons, by making minimal modifications. It can also be applied to different black hole solutions and to different heat capacities. One can also characterize and physically interpret the specific heats of black hole spacetimes beyond general relativity. It would be indeed interesting to check how the heat capacities, their divergences, and their physical interpretation in terms of a balance of pressures at the horizon compare with those of standard black holes when considering, for instance, black hole solutions coming from modified theories of gravity, effective black hole spacetimes coming from taking into account the effect of quantum fields, etc. On the other hand, the qualitatively similar behavior of black holes and other compact objects regarding the heat capacity could potentially reveal some aspects of the black hole microstates and internal structure during their thermal evolution under Hawking radiation. What happens around the divergence is particularly interesting in order to understand the origin of the thermodynamic phase transition usually associated with Davies points.

\bigskip

\section*{Acknowledgments}

PB and JAM acknowledge financial support from the Generalitat Valenciana through PROMETEO PROJECT CIPROM/2022/13. DFS is supported by the Grant No. PID2023-148373NB-I00 funded by MCIN /AEI /10.13039/501100011033 / FEDER, UE, by the Q-CAYLE Project funded by the Regional Government of Castilla y León (Junta de Castilla y León), and by the Ministry of Science and Innovation (MCIN) through the European Union funds NextGenerationEU (PRTR C17.I1).

\bibliography{biblio}
\end{document}